# Revealing Polytypism in 2D Boron Nitride with UV Photoluminescence

Jakub Iwański[1], Krzysztof P. Korona[1], Mateusz Tokarczyk[1], Grzegorz Kowalski[1], Aleksandra K. Dąbrowska[1], Piotr Tatarczak[1], Izabela Rogala[1], Marta Bilska[1,2], Maciej Wójcik[2], Sławomir Kret[2], Anna Reszka[2], Bogdan J. Kowalski[2], Song Li[3], Anton Pershin[3], Adam Gali[3,4,5], Johannes Binder[1], Andrzej Wysmołek[1]

[1]Faculty of Physics, University of Warsaw, Pasteura 5, 02-093 Warsaw, Poland
[2]Institute of Physics, Polish Academy of Sciences, Al. Lotników 32/46, 02-668 Warsaw, Poland
[3]HUN-REN Wigner Research Centre for Physics, P.O. Box 49, H-1525 Budapest, Hungary
[4]Department of Atomic Physics, Institute of Physics, Budapest University of Technology and Economics, Műegyetem rakpart 3., H-1111 Budapest, Hungary
[5]MTA-WFK Lendület "Momentum" Semiconductor Nanostructures Research Group, P.O. Box 49, H-1525 Budapest, Hungary

ABSTRACT. Boron nitride exhibits diverse crystal structures, predominantly a layered arrangement with strong intraplanar covalent bonds and weak interplanar van der Waals bonds. While commonly referred to as hexagonal BN (hBN), the sp$^2$-bonded BN atomic planes can also arrange in other configurations like Bernal (bBN) or rhombohedral (rBN) stacking orders. Variations in the orientation and translation of successive atomic layers lead to changes in crystal symmetry, potentially resulting in piezoelectric, pyroelectric or ferroelectric effects.

However, distinguishing between different polytypes using conventional methods like X-ray diffraction or Raman spectroscopy presents a significant challenge. In this work, we demonstrate that the optical response of the 4.1 eV defect can serve as an indicator of the polytype. To this end, we study BN samples grown by metalorganic vapor phase epitaxy (MOVPE), which contain different polytypes. The identification of the polytypes was achieved by X-ray diffraction and transmission electron microscopy. Photoluminescence and cathodoluminescence measurements with a high spatial resolution allowed for the deconvolution of the signal into two components from which we can extract a zero-phonon line (ZPL) at 4.096 eV (302.6 nm) for hBN and 4.143 eV (299.2 nm) for rBN. We performed calculations that enable us to identify the defect as a carbon dimer $C_BC_N$ ($C_2$) and show that the ZPL shift reflects differences in the crystal environment for different polytypes. Furthermore, we demonstrate that different polytypic composition ratios of hBN and rBN can be achieved by MOVPE, which could pave the way for future applications in large-area van der Waals heterostructures.

Boron nitride can crystallize in different phases including cubic and wurtzite phases that have $sp^3$-hybridized bonds and hexagonal or rhombohedral phases that are constructed by flat layers of atoms of $sp^2$ bonds, with a large separation between two layers joined by van der Waals bonds. However, at ambient conditions the $sp^2$ configurations are the stable phases [1–4]. Boron nitride in the $sp^2$-bonded hexagonal phase (hBN) is a very popular 2D material with insulating properties and is commonly used in various types of van der Waals heterostructures [5,6]. Color centers in boron nitride can serve as bright UV and visible single photon emitters [7,8]. Hexagonal boron nitride has an indirect band-gap, but shows very bright luminescence with nearly 100% efficiency [9–12]. Although hBN has a high resistivity and it is still unclear whether effective doping can be achieved, there is still hope for light emitting devices, since electroluminescence of pristine BN was recently demonstrated in van der Waals heterostructures [6].

The hBN polytype exhibits an AA' stacking sequence and is characterized by indirect bandgap [13–16]. However, the existence of other polytypes, such as Bernal BN (bBN) with AB stacking (where the second layer (B) is shifted along the m-axis in relation to layer A) and rhombohedral BN (rBN) with ABC stacking (with an additionally shifted C layer) was recently experimentally demonstrated [16–18]. The stacking sequence of atomic layers for different polytypes is plotted in Figure 1. All of these forms have different band structures, as predicted theoretically and observed experimentally via the near bandgap optical emission [3,13,14,19–21].

In this work, we demonstrate that defect-related optical emission in the UV-B spectral range is sensitive to the changes in the stacking sequence of atomic layers, suggesting that mid-gap photoluminescence can be used to identify the $sp^2$-BN polytype as previously suggested for bBN [22]. Specifically, we focus on the 4.1 eV color center emission, which has been extensively studied in the literature [10,23–27]. There is an ongoing debate regarding its origin, with various interpretations proposed: nitrogen vacancy [28], carbon ring [29], Stone-Wales defect [30]. However, the most widely accepted and probable interpretation is a carbon dimer ($C_2$) at boron and nitrogen sites, $C_B C_N$ [27,31–33]. Moreover, the existence of the carbon dimers in BN has been directly observed by high-resolution electron microscopy in an atom-by-atom analysis [34].

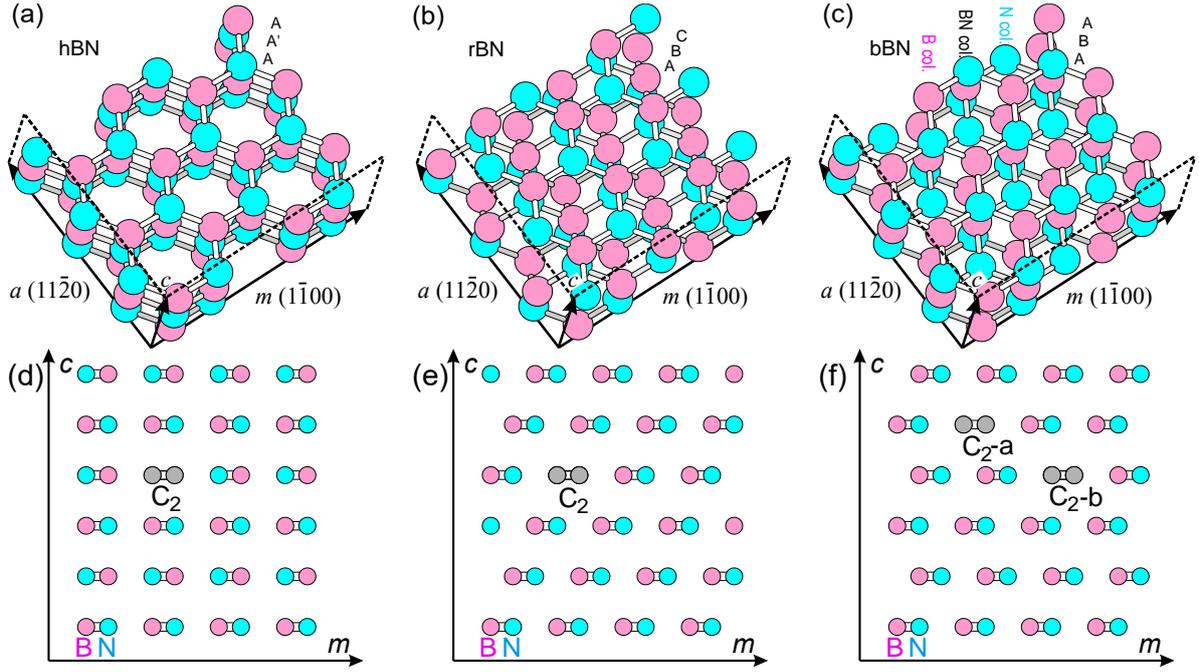

Figure 1: Different stacking orders of sp$^2$-BN layers resulting in different polytypes: a) hBN, b) rBN and c) bBN and arrangement of atoms surrounding carbon dimers $C_2$ ($C_B C_N$) in d) hBN, e) rBN, and f) bBN.

Boron nitride crystalizes in honeycomb layers stacked one onto another. The layers have hexagonal symmetry. The main crystallographic planes are the *c*-plane (plane normal to the layer surface), *a*-plane and *m*-plane. Figure 1 shows these planes. Their Miller and Miller-Bravais indices are: *c*: (001) and (0001), *a*: (110) and (11$\bar{2}$0), and *m*: (1$\bar{1}$0) and (1$\bar{1}$00). Here, we illustrate the concept of polytype identification using the carbon dimer $C_B C_N$ as an example. In sp$^2$-hybridized boron nitride, the B-N covalent bonds lay in *m*-plane: (1$\bar{1}$0), so a pair of carbon atoms that replaces boron and nitrogen atoms ($C_B C_N$) are positioned in the *m*-plane. Therefore, the most convenient way to analyze arrangement of atoms is in the *m*-plane cross-section view as depicted in Figs. 1d - f. Hexagonal boron nitride is formed by AA' stacking, where the A' layer is rotated by 60° what can be observed in Figs. 1a and 1d. Rhombohedral boron nitride is formed by consecutive layers shifted along the direction of B-N bonds (see Fig. 1b and 1e). Depending on the polytype, the surrounding of the incorporated carbon dimer changes. From the electronegativity of the atoms, it follows that boron atom is charged positively while nitrogen atom is charged negatively, so the B-N pair is an electric dipole. In rBN all such dipoles around the $C_BC_N$ pair are in the same direction, while in hBN the direction varies and dipoles compensate each other. The bBN stacking provides two options for $C_NC_B$ incorporation that are marked in Fig. 1f as $C_2$-a and $C_2$-b. In the case of the $C_2$-a dimer position, one C atom has two close boron neighbors in the upper and lower layer and the another C atom has no close neighbors. In the case of the $C_2$-b dimer, only one C atom is directly surrounded by two nitrogen atoms. These difference between the two positions may be the origin of two different emission lines [22,35].

Such small differences in the crystal structure of the material lead to significant changes in its properties. Different ordering induces changes in the band structure and causes a redshift of the energy gap value for rBN relative to hBN or a blueshift for bBN [3]. Additionally, the relative shifts and twists of subsequent atomic layers lead to a change in the symmetry group, which consequently induces piezoelectric, pyroelectric or ferroelectric effects [36]. Moreover, since rBN crystals have no inversion symmetry, they can display even-order nonlinear optical phenomena like the frequency doubling effect (second-harmonic generation) [37]. The ability to identify or produce the appropriate polytype opens up completely new possibilities for applications in two-dimensional heterostructures.

Due to the similar lattice constants, an unambiguous identification of polytypes for high-quality epitaxial layers or exfoliated flakes is particularly difficult when using conventional X-ray diffraction (XRD). On the other hand, the weak van der Waals interaction between atomic layers causes nearly no difference between polytypes in the energy of the most pronounced $E_{2g}^{high}$ (Raman active) or $E_{1u}$ (IR active) phonon modes, so Raman spectroscopy and other techniques that focus only on these phonons are not distinctive.

As we demonstrate below, the variations in the environment of the carbon dimer $C_BC_N$ notably impact the energy of the characteristic photoluminescence lines associated with defect emission. Consequently, the polytype can be readily and unequivocally identified.

Results and Discussion

One of the most common techniques for the structural characterization of crystals is X-ray diffraction (XRD). However, in contrast to powder-like samples for thin crystals and films there are serious limitations regarding observed peaks during the experiment. The strongest signal in the 2θ/ω scan is observed for (00*l*) Miller indices, and the other peaks cannot be observed because atomic planes are aligned parallel to the thick substrate. Since the distance between $sp^2$-BN atomic layers in different stacking orders is practically the same (d = 0.333 nm) [18], it is hard to experimentally distinguish between AA´, AB and ABC stacking sequence by observation of (002) and (003) peaks around $2\theta \approx 27°$ [18,20]. The main differences can be found for the peaks around $2\theta \approx 40° - 45°$. In hBN, in this angular range peaks related to (100) and (101) planes are expected, while in rBN, reflections from (101) and (102) planes should be present [18,20]. To access those peaks, the atomic planes in $sp^2$-BN have to be rotated relative to the substrate or the film has to be free-standing without substrate underneath. Therefore, to obtain a sample by metal organic vapor phase epitaxy (MOVPE) which allows to easily distinguish polytypes by XRD, we grew a sample under conditions which support rapid growth of misoriented BN flakes. The morphology of this sample, presented in the Supporting Information, indicates the presence of flakes misaligned at large angles with respect to the substrate. The random distribution of flake orientations allows us to treat the sample similar to a powder, providing access to diffraction conditions for (100), (101) and (102) atomic planes.

To meet the diffraction conditions for bulk hBN we conducted measurements for a free-standing crystal. To enable such measurements, we mounted a small bulk hBN crystal onto a substrate in a such a way that part of the crystal protrudes from the edge of the substrate allowing to measure the free-standing part of the hBN.

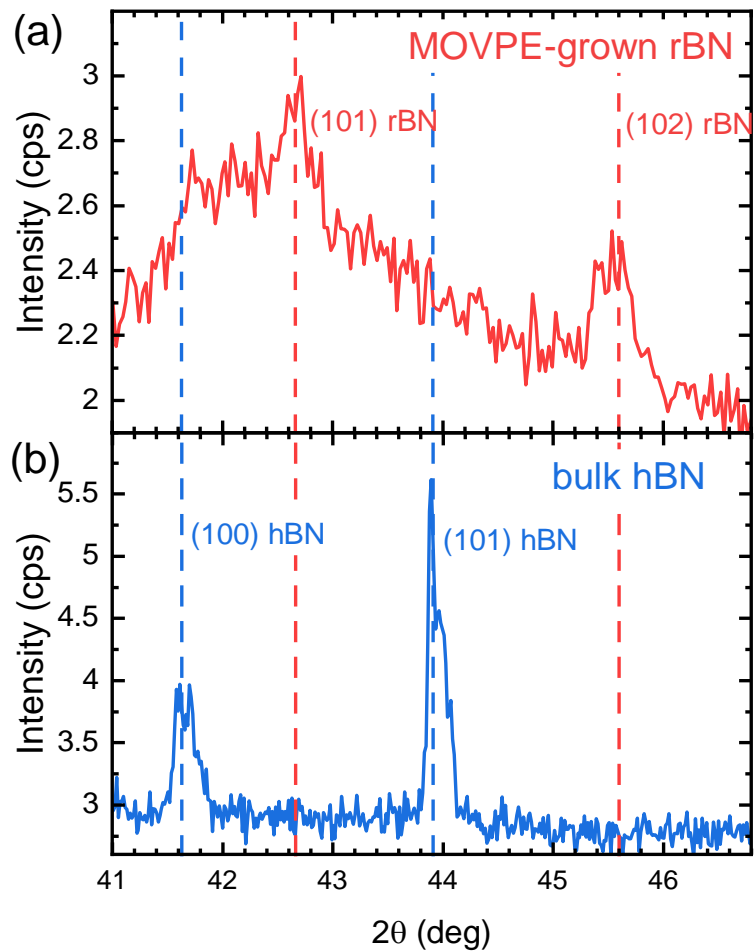

Figure 2: The X-ray diffraction 2θ/ω scans for a) MOVPE-grown rBN and b) free-standing, comercially available hBN bulk crystal measured as a reference. Blue and red dashed lines refer to the angular positions of hBN- and rBN-related XRD peaks, respectively.

In Fig. 2, we present the XRD results for the MOVPE-grown BN sample with randomly oriented flakes and commercially available bulk hBN. As can be seen in Fig. 2, for the MOVPE-grown sample only rBN-related (101) and (102) peaks were detected with no clear evidence of hBN. Broad asymmetric background signal originates most likely from some unconnected single atomic layers giving signal from (100) plane. Since the size of the X-ray beam is about 3x10 mm, the obtained signal is averaged over almost the whole sample. This leads to the conclusion that the sample is composed predominantly of rBN. In contrast to the MOVPE-grown sample, the 2θ/ω scan presented for bulk hBN crystal reveals two peaks at positions corresponding to (100) and (101) planes in hBN with no evidence of rBN inclusions. Throughout the further analysis, we use these two samples as a reference for rBN and hBN.

**UV-photoluminescence**. Pure boron nitride emits light mainly near its band edge in the UV-C range at about 6 eV, which is experimentally very challenging to measure [18]. However, in many samples also a defect-related mid-bandgap luminescence in the UV-B range is observed [7,10,23,24,26,27,38]. Very characteristic and intensively studied is the color center emission at about 4.1 eV that is composed of a few narrow lines, which is often interpreted as a carbon dimer ($C_2$), for which carbon occupies the closest-neighbor boron and nitrogen sites, 12-$C_BC_N$. [21,27,31–33]

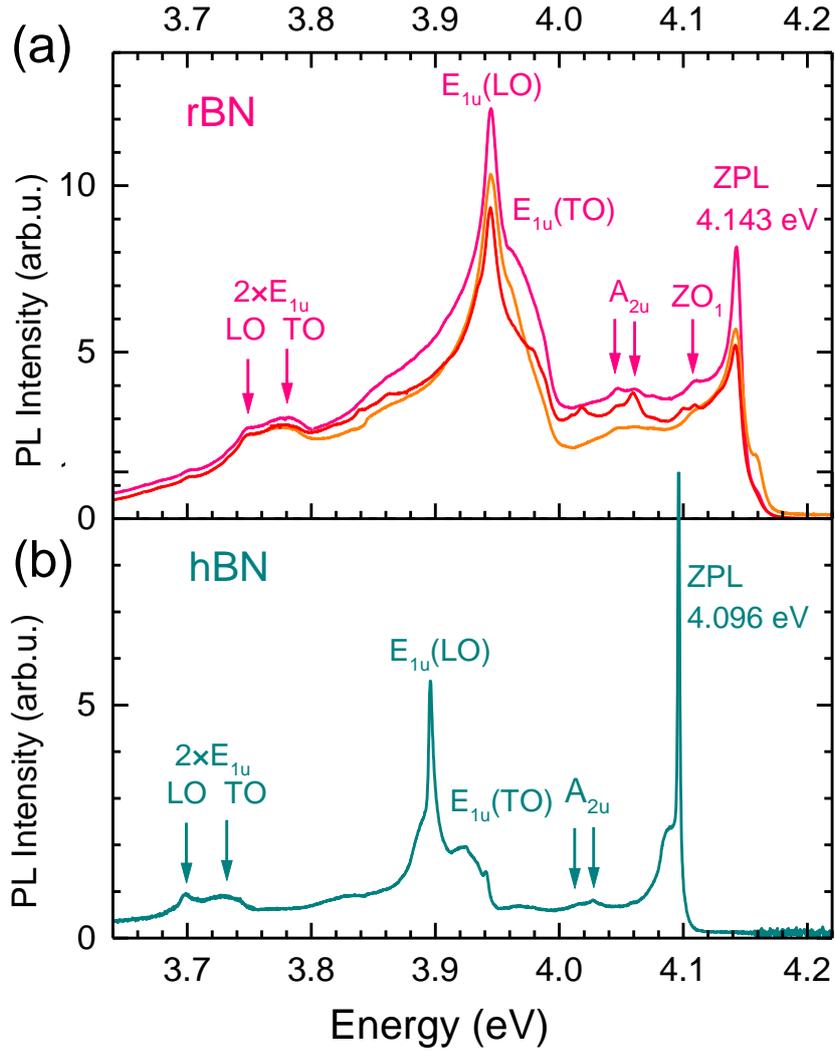

Figure 3: Photoluminescence spectra characteristic for a) three MOVPE-grown samples containing rBN and b) reference bulk hBN sample. Zero-phonon lines (ZPL) and selected phonon replicas characteristic for rBN and hBN are marked.

In fact, the photoluminescence (PL) spectra in the UV-B range observed for our samples exhibit strong zero-phonon lines (ZPL) with associated phonon replicas. As presented in Fig. 3, the spectra of MOVPE-grown rBN reference sample, for which clearly evident (101) and (102) XRD peaks were observed (Fig. 2), and two samples of similar spectra. The ZPL for rBN is observed at 4.143 eV (299.2 nm) which is blueshifted by $\Delta E_{ZPL} = 47$ meV in respect to hBN ZPL at 4.096 eV (302.6 nm). Both ZPLs are followed by similar shift of dominant phonon mode at ~200 meV. This shift is most probably a result of the different local surrounding of a defect in rBN and hBN (see Fig. 1). It is worth to notice that in the case of bBN, two ZPLs were already reported at 4.14 eV and 4.16 eV [22]. The coincidence of the 4.14 eV line in bBN with the ZPL in rBN is not unexpected. As depicted in Figure 1, the defect surroundings in rBN and bBN exhibit similarities. In some of our MOVPE-grown samples, a small peak at 4.16 eV is also observed, which suggests the presence of bBN inclusions in our epitaxial layer. Usually this contribution is very small, therefore, it does not influence our further interpretations.

The ZPLs for both types of samples are quite narrow. In the case of the reference sample (hBN), the ZPL is symmetric and its full width at half maximum (FWHM) is 1.5 meV (0.12 nm). In the case of the 299.2-nm line (assumed as rBN), the ZPL is asymmetric. At the high energy side it has a steep slope related to a Gaussian curve with an FWHM of 8 meV (0.7 nm). Such a narrow width suggests a well-ordered crystal structure.

At the low energy side, the ZPLs are broadened most likely by interactions with acoustic phonons, which is also supported by our first principles calculations (see Supplementary Information for details). The symmetric broadening is probably caused by inhomogeneities of the sample structure, especially by a turbostratic twisting of atomic layers.

As presented in the Supplementary Information, the lifetime of PL lines is of the order of 1 ns, therefore the homogenous broadening is negligible (of the order of a few μeV). Since the broadening is much lower than the energy shift between ZPLs in the two materials ($\Delta E_{ZPL}$ = 47 meV), the $\Delta E_{ZPL}$ must be due to a regular, well-defined change of the defect structure or ordering of the crystal such as a change of the polytype. The interaction between the covalently bonded atoms involves energies of many eV and the density functional theory calculation presented in Ref. [31] shows that a modification of the defect structure (adding or removing neighboring atoms) changes the electronic states of defects significantly. Taking this into account, the $\Delta E_{ZPL}$ shift is much smaller in comparison to the energy expected for changes of covalent bonds of the defect. However, the order of magnitude of the $\Delta E_{ZPL}$ shift is right when considering the influence of translations between subsequent atomic layers in different polytypes.

It can be noticed that the PL background observed for MOVPE-grown samples is higher than for bulk BN. We believe that the broad background signal is most likely originating from a disorientation of atomic planes. Such a broad emission was reported in ref. [27] as D330 band.

The PL spectra for reference hBN and rBN presented in Figure 3 exhibit a rich structure of phonon replicas. The most intensive replicas are related to the $E_{1u}$ mode. For the $E_{1u}$ mode the atoms vibrate only in the *c*-plane (001), which is parallel to the $sp^2$-BN layer [5,39]. We do not expect that a difference in stacking would significantly influence such phonons. In fact, the $E_{1u}(LO)$ replica has nearly the same energy 198 meV and 200 meV in rBN and hBN, respectively. Similar observations are noted for $E_{1u}(TO)$ and two-$E_{1u}(LO)$-phonon replicas.

However, the situation changes for the $A_{2u}$ mode, which represents vibrations perpendicular to the *c*-plane [19,39,40]. Therefore, the alteration in stacking order is anticipated to significantly impact these modes, thus we experimentally observe shifts in their energies. The peaks in the $A_{2u}$ region, indicated by arrows, can be associated with phonons of energies 82 meV and 96 meV in rBN and 70 and 80 meV in hBN.

Time resolved luminescence shows that the lifetimes are of the order of 1 ns. For the hBN reference sample from Fig. 3b, the measured lifetimes are $\tau = (0.95\pm0.05)$ ns for both ZPL and $E_{1u}(LO)$ phonon replica. In the case of MOVPE-grown samples a variation of lifetimes ranging from 0.7 to 0.8 ns was observed for the narrow-line samples with a pure 4.14-eV line (rBN). Generally, even for samples with a mixed polytypic composition, which show spectral lines both at 4.14-eV (rBN) and 4.10-eV (hBN), lifetimes are above 0.5 ns. In other semiconductors, a large number of defects usually causes fast nonradiative recombination. The fact that we do not observe large differences in PL emission lifetimes between high-quality bulk hBN and mixed-polytype epitaxial BN samples can be understood by small overlapping of wave functions of different color centers and other defects. Both the thermal escape and tunneling are suppressed by high barriers, so the color centers are not substantially sensitive to the presence of other defects. Similar lifetimes were also reported in literature [10,27].

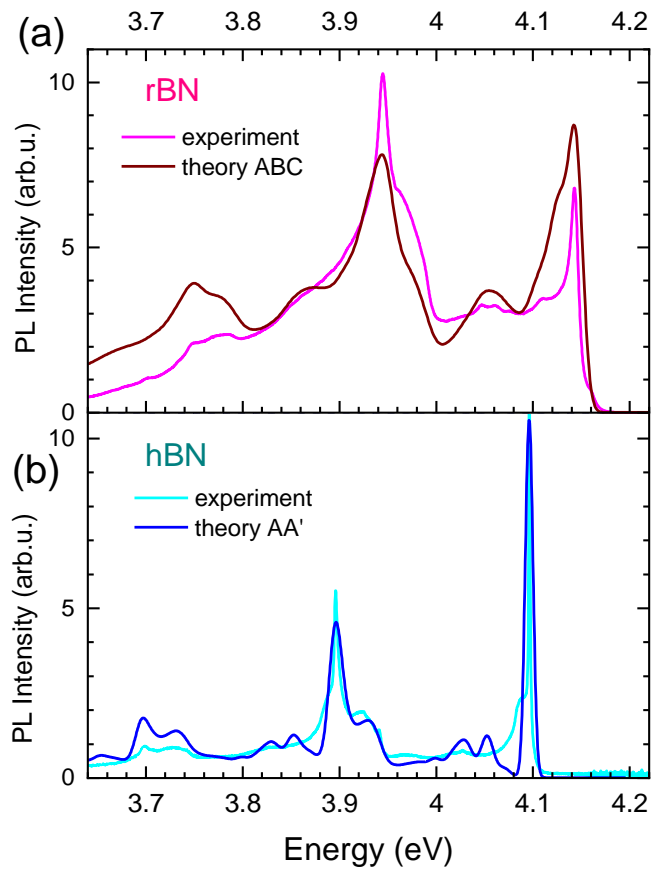

Figure 4: Comparison of calculated emission spectra (dark lines) with experimental PL spectra (light lines) for a) rBN and b) hBN.

To identify the defect responsible for the observed photoluminescence, we compared experimental results with theoretical calculations concerning emission within this spectral range [35]. Our analysis pointed towards the carbon dimer $C_BC_N$ as the most promising candidate. Subsequently, we performed spin-polarized density functional theory (DFT) calculations to determine the emission spectra of $C_BC_N$ defects in $sp^2$-BN with various stacking orders. To this end, we developed a new multiconfiguration ΔSCF method that mimics the spin-adapted wave functions in the optical excited state force calculations (see Supplementary Information). We applied it to a periodic model to simulate an infinite number of layers. The obtained emission lines were modelled taking into account an "effective" broadening effect, by assigning Gaussian functions to each active mode. The broadening may be due to a specific modification in a stacking pattern, such as turbostratic twisting. The resulting spectra for the $C_BC_N$ defect show a very good agreement with our experiments for both rBN and hBN, as presented in Figure 4. The ZPL of the carbon dimer emission energy was obtained equal to 4.21 eV and 4.09 in rBN and hBN, respectively. This accuracy is excellent especially considering that the standard ΔSCF method typically underestimates the ZPL energies of $C_BC_N$ by ~0.5 eV [32]. While a correction term has been applied to the excitation energy to address this discrepancy [29,32], it has not been applied to the forces. We found it crucial to accurately calculate the forces as well in order to closely match the experimental spectra of $C_BC_N$ in rBN, as demonstrated in Supplementary Information Fig. S3.

The interesting comparison between the calculated photoluminescence spectra lies in the phonon replicas. The main difference in the carbon dimer spectra for hBN and rBN can be observed in the range related to the phonons of $A_{2u}$ symmetry which are out-of-plane phonon modes, as well as in the low-energy acoustic phonon range (see Supplementary Information Fig. S4). The $A_{2u}$ phonon is experimentally identified in rBN at 92.6 meV. As expected, the contribution of the $A_{2u}$ phonon in hBN is not observed, given that the $C_BC_N$ carbon defect incorporates within the $sp^2$-plane. However, a small contribution of the $A_{2u}$ phonon is visible in rBN, attributed to the out-of-plane distortion. This three-layer feature is also briefly mentioned in Ref. [5]. Another notable difference between hBN and rBN is the emergence of low-energy out-of-plane phonons at ~20 meV. This effect is directly observable in our experiment as an asymmetric broadening of ZPL. Interestingly, no substantial spectral broadening has been observed for the defect emission in bBN [22]. This is due the $C_{2v}$ symmetry of $C_BC_N$ defect in hBN and bBN, which prohibits the coupling of its $A_1$ electronic wavefunctions to $A_2$ phonons. In contrast, in rBN where the defect exhibits $C_{1h}$ symmetry, this constrain is lifted, leading to spectral broadening as an indication of the absence of $C_{2v}$ symmetry, or in other words, an asymmetric arrangement of neighboring layers. The cumulative effect of the out-of-plane phonons in rBN is manifested as an out-of-plane distortion in the excited state of ~3 degree. Moreover, as outlined in this paragraph and depicted in Figure 4, the experimental spectra of $sp^2$-BN are in great agreement with the theoretically calculated spectra of the $C_BC_N$ defect. Hence, we attribute the observed spectra to the $C_BC_N$ defect. This interpretation, previously proposed, aligns with the omnipresence of carbon in MOVPE technology, yielding luminescence across various spectral ranges [27,41]. Furthermore, $C_BC_N$ represents the smallest possible carbon cluster and is inherently present in triethylboron, the boron precursor utilized during the growth process.

**Application of UV luminescence analysis.** As presented before, using MOVPE it is possible to grow rBN layers consisting of randomly oriented flakes. However, for most of the applications flat, high-quality $sp^2$-BN is required. This raises the question, whether is it possible to control polytypic composition of high-quality material by adjusting MOVPE growth conditions. To answer this question, we grew a series of samples using slightly different conditions, and we choose a set of three characteristic samples ($S_1$, $S_2$, $S_3$). More details about the growth process can be found in Experimental methods and characterization results in Supporting Information.

In Figure 5, we present typical high-resolution transmission electron microscopy (HRTEM) images for samples $S_1$, $S_2$, and $S_3$. Images of a larger area are presented in the Supporting

Information. It is evident that all three polytypes are present in our samples. However, an analysis of multiple images for each sample suggests variations in polytypic composition among them. Sample $S_1$ (Fig. 5a) exhibits the highest contribution from rBN, with other polytypes being in the minority. Conversely, sample $S_3$ (Fig. 5c) displays the most significant amount of hBN, although other polytypes also contribute significantly. Sample $S_2$ demonstrates the least polytypic homogeneity, with all polytypes easily discernible.

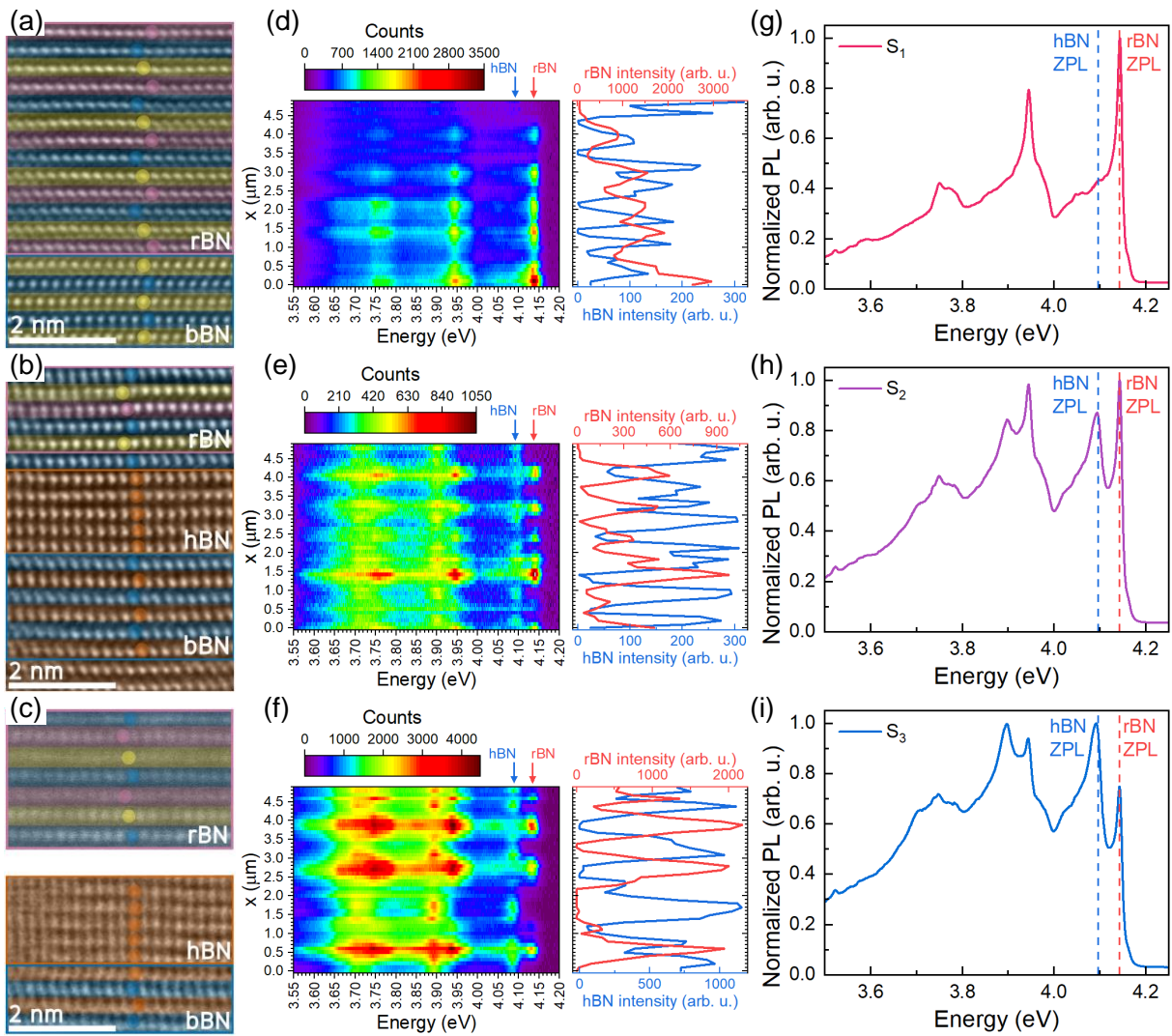

Figure 5: a) - c) Exemplary transmission electron microscopy (TEM) pictures for the samples $S_1$, $S_2$, $S_3$. The overlaid colors depict atomic layers which are aligned similarly. The colored circles are guide to the eyes. d) - f) Hyperspectral line-scans of cathodoluminescence collected

along a 5 µm-long line on the sample. Curves in the panel on the right side of the map indicate the intensity of the spectral components which correspond to rBN (red) and hBN (blue). g) – i) Macro-photoluminescence spectra measured at 8 K for epitaxial samples with different polytypic composition. Arrows in d) – f) and dashed lines in g) – i) indicate the position of zero-phonon lines for rBN- and hBN-related spectra.

Since HRTEM only provide an insight into the local crystal structure, we also performed top-view cathodoluminescence (CL) hyperspectral imaging along randomly chosen 5 µm-long lines. As can be seen in Figure 5d-f, luminescence is detected only in some segments of the line suggesting that the optical response of the material is indeed defect-related, since it is possible to resolve emission from isolated defects. Furthermore, intensity variations of the ZPL can be observed as well as changes of the ratio between the spectral components of rBN and hBN. To make the analysis more quantitative we deconvoluted the spectra into five components: rBN-related spectrum, hBN-related spectrum, and three broad Gaussian curves: at the maxima of the ZPL and at two of the strongest phonon replicas. Such background components can be physical explained as a twist of subsequent atomic layers which leads to the broadening of spectral lines as presented in Ref. [35]. By fitting the intensity of all five components, we could extract the contribution of rBN and hBN for each collected spectrum. This result can be found in the right panels of Fig. 5d-f and in the form of a video in the supplementary information. It is noteworthy that the intensity ratio of spectral components does not necessarily reflect the polytypic composition ratio, as the optical response and efficiency of defect creation may vary among different polytypes. The details of the fitting procedure can be found in the Supporting Information. While the $S_1$ spectrum is dominated by rBN component (Fig. 5d), CL spectrum of the sample $S_2$ comprises a mixture of components associated with both rBN and hBN (Fig. 5e). The most intense hBN-related component we observed in sample $S_3$, for which the widest areas of hBN were found in HRTEM.

To determine the average polytypic composition in each sample, we conducted macro-photoluminescence experiments with a spot size of approximately 50 µm. Consistent with CL findings, we observe varying intensity ratios of ZPLs at 4.14 eV and 4.10 eV, indicating differences in polytypic composition among the samples. As anticipated from TEM and CL analyses, sample $S_1$ exhibits a prominent peak at 4.14 eV with characteristic phonon replicas (Fig. 5g) suggesting that the sample is indeed dominantly rBN. In the PL spectra of samples $S_2$ and $S_3$ (Fig. 5f-h), distinct and well-resolved ZPLs are observed at both energies: 4.14 eV and 4.10 eV. The changing intensity ratio of these peaks suggests a varying composition with a higher fraction of hBN in sample $S_3$. Furthermore, collecting macro-PL spectra over a wide spatial area allowed us to detect a very subtle signal at 4.16 eV associated with bBN, consistent with the presence of bBN observed in HRTEM images.

By comparing these three samples, we demonstrate that it is possible to obtain a different composition ratio between hBN and rBN by using wafer-scale MOVPE growth technology. Although more work is needed to demonstrate the precise control over the growth of a desired polytype in flat, epitaxial $sp^2$-BN layers, the presented results suggest that this goal may be achievable by MOVPE growth of $sp^2$-BN on sapphire, which could open up the way for scalable growth of different BN polytypes using a well-established, industrial growth method.

## CONCLUSIONS

In summary, we have demonstrated that different $sp^2$-BN polytypes can be grown by MOVPE and we present a facile way to distinguish between these polytypes. We have shown that the characteristic, impurity-related UV photoluminescence with a ZPL at about 4.1 eV is influenced by the structural polytype of the material. The observed lines are quite narrow and shift significantly with a change of stacking order. We observed emission energies for ZPLs at 4.096 eV and 4.143 eV, which, according to X-ray diffraction results, can be ascribed to defect emission in hBN and rBN, respectively. The energy difference of $\Delta E_{ZPL} = 47$ meV is much larger than the line widths (FWHM about 8 meV), which allows us to clearly distinguish between $sp^2$-BN polytypes. Cathodoluminescence measurements with sub-micrometer resolution facilitate a deconvolution of the overall spectra to estimate the local ratio of hBN/rBN - related peaks for samples consisting of a mixture of polytypes. To support our experimental findings, we have performed first principles DFT calculations that show that all characteristic features of the spectra can be reproduced when considering a $C_B C_N$ ($C_2$-dimer) defect embedded in different $sp^2$-BN polytypes. Although we utilized the $C_B C_N$ defect emission to sense the stacking sequence of $sp^2$-BN, it is likely that other defects are also sensitive to polytypism.

While in the case of bulk $sp^2$-BN crystal growth, the most commonly obtained polytype is hBN, for epitaxial growth, it is possible to obtain both hBN and rBN. We demonstrate that by MOPVE one can grow high-quality, flat BN layers with different polytypic composition, as evidenced by high-resolution transmission electron microscopy, cathodoluminescence and macro photoluminescence, which are measurements techniques that cover a large range of length scales. Therefore, our results not only present a facile way to distinguish between different $sp^2$-BN polytypes, but also constitutes a first step towards the deterministic control over the growth of $sp^2$-BN polytypes on the wafer-scale. By using MOVPE, which is a well-established industrial method, it may become possible to make use of piezoelectric or ferroelectric properties in large-area two-dimensional heterostructures.

Experimental methods

**Samples** The boron nitride layers on sapphire substrates were grown using Metalorganic Vapor Phase Epitaxy (MOVPE) using an Aixtron CCS 3x2" MOVPE system designed for the growth of nitride compounds. The epitaxial rBN samples were grown on 70 nm-thick AlN template at T = 1025 - 1100°C using Continuous Flow Growth (CFG) mode [42,43]. The temperature was controlled by ARGUS optical pyrometer. High-quality samples were grown using Two-stage Epitaxy growth protocol on 2-inch sapphire substrates with 0.6° towards *m*-plane misorientation angle [44,45]. CFG buffer layer was grown for 10 min at 1300 °C for each sample. After this step the temperature was ramped up to 1400 °C and where the Flow Modulation Epitaxy stage occurred. The ramping temperature time varied between samples: $S_1$ – 15 min, $S_2$ – 8 min, $S_3$ – 5 min. The reference bulk hBN sample was provided by HQ Graphene company.

**X-ray diffraction** XRD 2θ/ω scans were measured with a Panalitical X'Pert setup with standard CuKα X-ray source. The beam was formed by the symmetrical Bragg–Brentano optics. To obtain rocking curves of (101) and (102) reflections for the reference MOVPE-grown rBN the offset of 7° in respect to diffraction conditions for sapphire substrate was established to avoid strong signal from crystalline substrate and observe diffraction on misoriented rBN crystallites. To observe diffraction on hBN bulk flake it was sticked to the substrate in such a way that significant size of the flake was out of plane of the substrate.

**First principles simulations** The first principles calculations were carried out within the Vienna ab initio simulation package (VASP) code [46,47], using a plane wave basis and projector augmented wave (PAW) potentials [48,49]. The defects were embedded in a 9 × 9 bilayer supercell model to avoid the interactions of defect with its periodic images. The Γ-point approximation was used for Brillouin-zone sampling. We used the DFT-D3 method of Grimme [50] for dispersion correction of the vdW interaction. The hybrid density functional of Heyd, Scuseria, and Ernzerhof [51] was used to optimize the geometries and electronic properties with modified mixing parameter $\alpha = 0.32$ of Fock exchange. The cutoff energy of plane-wave is 450 eV and all ions were allowed the relax until the Hellman-Feynman forces were less than 0.01 eVÅ$^{-1}$. The excited states were calculated with a new multiconfiguration $\Delta$SCF method, presented in Supplementary Information.

**Transmission electron microscopy** The electron transparent samples for cross-section imaging were prepared using focused gallium ion beam in a FEI Helios NanoLab 600 system (FIB). HRTEM investigations were performed using an image corrected Titan Cubed 80-300 microscope operating at 300 kV.

**Photoluminescence** For photoluminescence (PL) spectroscopy, the samples were excited with light generated as third harmonic (in the range 4.2 - 5.1 eV) of a Ti:Sapphire laser. Two BBO crystals were used to obtain the short-wavelength light. A Schwarzschild mirror objective was used for micro-PL spectroscopy. The spectra were recorded with an Acton 300-mm spectrometer and a Hamamatsu CCD camera. The time-resolved PL (TRPL) was measured using a Hamamatsu streak camera with quartz optics (transmission up to 200 nm). Macro photoluminescence spectra were acquired from high-quality samples using the 4th harmonic of a continuous wave mode-locked Ti:Sapphire laser (196 nm, 50 μm) at 8 K. Spectra were recorded with a Czerny-Turner monochromator ($f = 300$ mm) featuring a 1800 grooves/mm grating and a back-illuminated CCD camera (Andor Newton 920).

**Cathodoluminescence** To study local optical properties Hitachi SU-70 scanning electron microscope equipped with Gatan MonoCL3 cathodoluminescence (CL) spectrometer was used.

To avoid samples electron charging, samples were delaminated from sapphire substrate and transferred onto p-doped Si/SiO$_2$ substrate [52]. CL line-scans collected using hyperspectral mode with the CCD camera were used to record the local energy changes of the spectral features in a selected area on the sample surface. Series of spectra were collected along the specified 5 um long lines with the step of about 100 nm and collecting time per pixel – 40 s. All the collected spectra were measured using accelerating voltage of 5 kV, the beam current of 3 nA (the parameters were optimized for the best intensity to spatial resolution) at temperature of 5 K and with use of diffraction grating with 1200 l/mm optimized at 300 nm.

ACKNOWLEDGMENT

We thank Prof. Guillaume Cassabois and Prof. Bernard Gil for the possibility to measure the macro-PL spectra. This work was supported by the National Science Centre, Poland under decisions 2020/39/D/ST7/02811, 2022/45/N/ST5/03396, 2022/45/N/ST7/03355 and 2022/47/B/ST5/03314. A.G. acknowledges the support from the National Research, Development and Innovation Office of Hungary (NKFIH) in Hungary for the Quantum Information National Laboratory (Grant No. 2022-2.1.1-NL-2022-0000) and the EU HE project SPINUS (Grant No. 101135699). A part of the calculations was performed using the KIFÜ high-performance computation units. A.P. acknowledges the financial support of Janos Bolyai Research Fellowship of the Hungarian Academy of Sciences.

# Supporting Information:

# Revealing Polytypism in 2D Boron Nitride with UV Photoluminescence

Jakub Iwański[1], Krzysztof P. Korona[1], Mateusz Tokarczyk[1], Grzegorz Kowalski[1], Aleksandra K. Dąbrowska[1], Piotr Tatarczak[1], Izabela Rogala[1], Marta Bilska[1,2], Maciej Wójcik[2], Sławomir Kret[2], Anna Reszka[2], Bogdan J. Kowalski[2], Song Li[3], Anton Pershin[3], Adam Gali[3,4,5], Johannes Binder[1], Andrzej Wysmołek[1]

[1]*Faculty of Physics, University of Warsaw, Pasteura 5, 02-093 Warsaw, Poland*
[2]*Institute of Physics, Polish Academy of Sciences, Al. Lotników 32/46, 02-668 Warsaw, Poland*
[3]*HUN-REN Wigner Research Centre for Physics, P.O. Box 49, H-1525 Budapest, Hungary*
[4]*Department of Atomic Physics, Institute of Physics, Budapest University of Technology and Economics, Műegyetem rakpart 3., H-1111 Budapest, Hungary*
[5]*MTA-WFK Lendület "Momentum" Semiconductor Nanostructures Research Group, P.O. Box 49, H-1525 Budapest, Hungary*

Morphology of MOVPE-grown reference rBN

The scanning electron microscopy (SEM) images depicting the sample morphology were acquired using a FEI Helios NanoLab 600 system. To be able to detect an X-ray diffraction signal at about $2\theta \approx 40° - 45°$ the atomic planes of $sp^2$-BN have to be misaligned with respect to the substrate surface. The morphology of the reference MOVPE-grown rBN sample, comprising protruding flakes, is illustrated in Figure S1. The XRD result presented in Figure 2 in the main text was obtained for this sample.

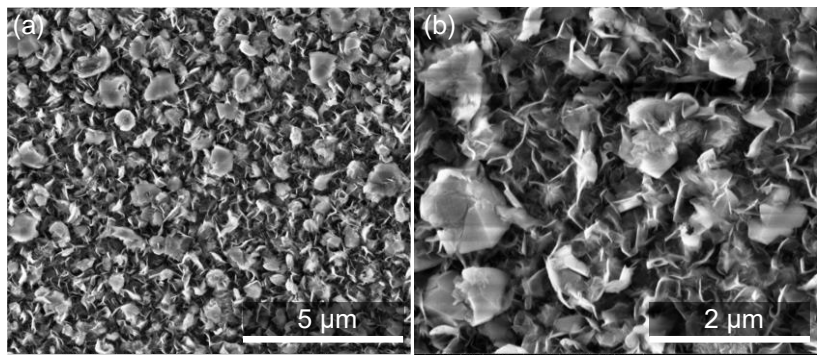

Figure S1: Scanning electron microscopy image of reference MOVPE-grown rBN.

## Time-resolved photoluminescence

Figure S2a-c displays time-resolved photoluminescence spectra for three sample classes: MOVPE-grown rBN with a single ZPL at 4.14 eV (Fig. S2a), which is similar to the reference presented in the main text (Fig. 2), MOVPE-grown mixed polytypic sample with two ZPLs at 4.14 eV and 4.10 eV (Fig. S2b), which is similar to the sample $S_2$ from the main text (Fig. 5), and bulk hBN purchased from HQ Graphene with a single ZPL at 4.10 eV (Fig. S2c). The observed ZPL lifetimes for each sample class are just below 1 ns.

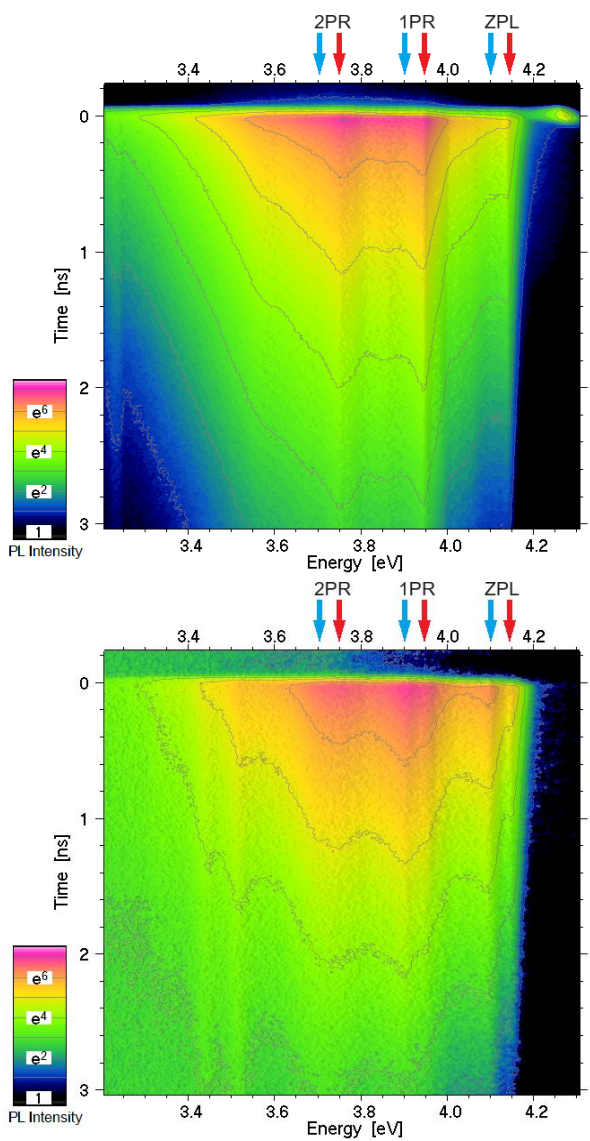

Figure S2a Time resolved PL (TRPL) spectrum of a sample with peaks related only to rBN.
ZPL – zero phonon line, 4.14 eV.
1PR – first phonon replica 3.95 eV.
2PR – second phonon replica 3.75 eV.
TRPL is presented as contour map. The isolines are equally spaced in logarithmic scale (e-times), so the distance between two contour lines along time axis gives decay time of the signal.
The lifetimes are 0.80(3) ns for all lines.

Figure S2b TRPL spectrum of a sample with peaks related to both hBN (blue arrows) and rBN (red arrows).
For rBN related lines, the lifetimes were found to be 0.75(5) ns and 0.78(4) ns for the ZPL and 1PR, respectively.
For hBN related lines, the lifetimes were found to be 0.78(4) ns and 0.80(4) ns for the ZPL and 1PR, respectively.
Two additional lines can be observed at 3.45 eV and 3.52 eV. Their lifetimes are above 2 ns.

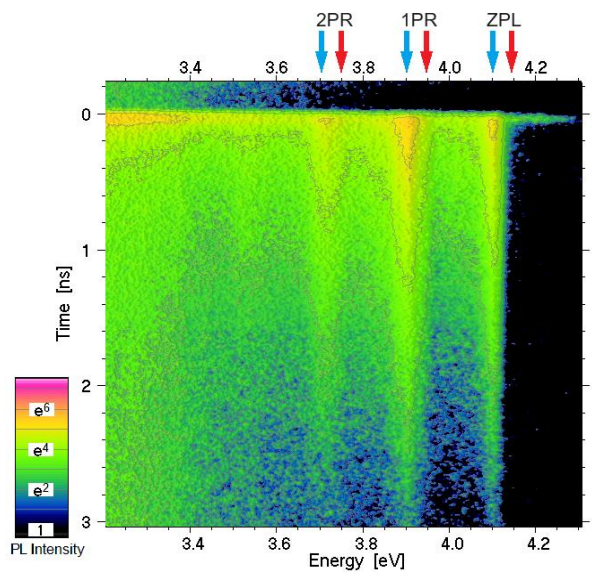

Figure S2c TRPL spectrum of the reference bulk sample with peaks related only to hBN.

ZPL – zero phonon line, 4.10 eV.
1PR – first phonon replica 3.90 eV.
2PR – second phonon replica 3.70 eV.
For these three lines, the lifetimes are 0.95(5) ns.

# Spin-polarized density functional theory (DFT) calculations

## *1. Calculations of excited state geometries.*

To accurately model the photoluminescence spectra of $C_BC_N$ defects in various stacking sequences of boron nitride, we have developed a new method to calculate the forces for multiconfiguration (MC) singlet excited states within the ΔSCF framework; further referred to as MC-ΔSCF. To this end, we first analyzed the multiconfiguration wavefunctions of the lowest excitations in both singlet ($^1\Psi$) and triplet ($^3\Psi_0$, $^3\Psi_+$, and $^3\Psi_-$) manifolds, given as follows:

$$^1\Psi = 2^{-1/2}(|\uparrow\downarrow\rangle - |\downarrow\uparrow\rangle), \quad (1a)$$

$$^3\Psi_0 = 2^{-1/2}(|\uparrow\downarrow\rangle + |\downarrow\uparrow\rangle), \quad (1b)$$

$$^3\Psi_+ = |\uparrow\uparrow\rangle, \quad (1c)$$

$$^3\Psi_- = |\downarrow\downarrow\rangle. \quad (1d)$$

Here, we consider an electron transition between the highest occupied and lowest unoccupied defect orbitals, specifying the spin projections by arrows. To proceed, the respective one-electron configurations are expressed in a basis of the multiconfiguration wavefunctions as follows:

$$|\uparrow\downarrow\rangle = 2^{-1/2}(^1\Psi + ^3\Psi_0), \quad (2a)$$

$$|\downarrow\uparrow\rangle = 2^{-1/2}(^1\Psi - ^3\Psi_0), \quad (2b)$$

$$|\uparrow\uparrow\rangle = ^3\Psi_+, \quad (2c)$$

$$|\downarrow\downarrow\rangle = ^3\Psi_-. \quad (2d)$$

The energy of each state is calculated as an expectation value of Hamiltonian, $H$, as $\langle \Psi | H | \Psi \rangle$. The energies of single configurations are directly accessible by the ΔSCF calculations. Given that the multiconfiguration wavefunctions are orthogonal, this yields the following relationships:

$$\langle \uparrow\downarrow | H | \uparrow\downarrow \rangle = 2^{-1}(E_S + E_T) = E_S^{\Delta SCF}, \quad (3a)$$

$$\langle \downarrow\uparrow | H | \downarrow\uparrow \rangle = 2^{-1}(E_S + E_T) = E_S^{\Delta SCF}, \quad (3b)$$

$$\langle \uparrow\uparrow | H | \uparrow\uparrow \rangle = E_T = E_T^{\Delta SCF}, \quad (3c)$$

$$\langle \downarrow\downarrow | H | \downarrow\downarrow \rangle = E_T = E_T^{\Delta SCF}. \quad (3d)$$

Here, $E_S$ and $E_T$ represent the energies of multiconfiguration singlet and triplet states, respectively. The superscript ΔSCF denotes the energies obtained by the ΔSCF method. Eqs. (3) demonstrate that the ΔSCF method can be used directly to calculate the triplet energy, but not the singlet state. Instead, the energy of the multiconfiguration singlet state is calculated using eqs. (3a) and (3c) as follows:

$$E_S = 2E_S^{\Delta SCF} - E_T^{\Delta SCF} = E_S^{\Delta SCF} + (E_S^{\Delta SCF} - E_T^{\Delta SCF}). \qquad (4)$$

Eq. (4), also known as a triplet correction scheme [1,2], demonstrates that the energy of an excited singlet state can be obtained by the ΔSCF method by combining the energies of individual configurations, providing a basis to the MC-ΔSCF method.

Furthermore, to calculate the forces, $F$, with the MC-ΔSCF method, we write the derivatives of energies in eq. (4) with respect to the nuclear coordinates as follows:

$$F_S = -\nabla E_S = -2\nabla E_S^{\Delta SCF} + \nabla E_T^{\Delta SCF}$$
$$= 2F_S^{\Delta SCF} - F_T^{\Delta SCF}. \qquad (5)$$

The analytic forces $F_S^{\Delta SCF}$ and $F_T^{\Delta SCF}$ are directly available from the ΔSCF calculations. Hence, the relaxation of geometry in the excited singlet state involves performing two force calculations at each geometry, followed by adjusting the atomic positions based on the chosen optimization protocol. For the purposes of this study, a steepest descent method was employed for its simplicity. More specifically, the atomic positions, $G_i$, were iteratively updated as follows:

$$G_{i+1} = G_i + k \cdot F_S, \qquad (6)$$

using a step constant, $k$, of 0.015 in a Cartesian basis. The procedure is repeated until reaching the convergence thresholds for the energy and forces. It is important to highlight that this methodology is readily extendable to different spin multiplicities and even to correlated wavefunctions, with coefficients obtainable through advanced techniques like time-dependent DFT.

## 2. Simulated spectra of $C_BC_N$ defect in hBN and rBN.

Furthermore, we apply our new method to simulate the photoluminescence spectra of $C_BC_N$ defect in the two polymorphs of boron nitride. Figure S3 compares the results obtained with our MC-ΔSCF methodology with those from the standard ΔSCF. Note that the standard ΔSCF method was applied to minimize the $E_S^{\Delta SCF}$ energy, defined in eq. (3a). Our low temperature measurements were used as reference. Figures S3b and S3d show a great agreement between our theoretical predictions and experiment in the AA' polytype (regular hBN). This observation supports our assignment of the 4.1-eV PL spectra to the $C_BC_N$ defect. Notably, both theory approaches show similar PL spectra. From a technical perspective, this consistency validates our methodology, particularly given that the geometry derived from the standard ΔSCF was obtained using the conjugate-gradient method. Furthermore, our new method demonstrates a better agreement with the reference data, particularly concerning a shoulder at ~3.93 eV. In contrast to the AA' polytype, the standard ΔSCF method fails to accurately reproduce the PL sideband in rBN, as shown in Figure S3c. However, our MC-ΔSCF method provides a very good agreement with the experimental results, closely replicating both the intensities and frequencies of active modes.

As discussed in the main text, the zero-phonon line of the 4.1-eV emission signal exhibits a blue shift of ~50 meV from hBN and rBN. Additionally, our calculations demonstrate a substantial alteration in the Huang-Rhys factors, increasing from 2 in hBN to 2.8 in rBN. This observed difference is attributed to a symmetry reduction of the $C_BC_N$ defect from $C_{2v}$ to $C_{1h}$, as previously discussed in ref. [1]. The change in the Huang-Rhys factors is mainly associated with an out-of-plane distortion of the $C_BC_N$ defect by ~3 degree in the excited state. The partial Huang-Rhys factors are shown in Figure S4a. Both plots for the $C_BC_N$ defect in hBN and rBN exhibit similar phonon activity at around 200 meV. The difference in the photoluminescence spectra is due to the low energy modes at ~20 meV, with two of them displayed in Figure S4b. Notably, these out-of-plane modes break the $C_{2v}$ symmetry and therefore do not occur in the AA' stacking. Importantly, these phonons and their overtones directly contribute to the spectral broadening, which along with the zero-phonon line shift, acts as a fingerprint of the polytype change.

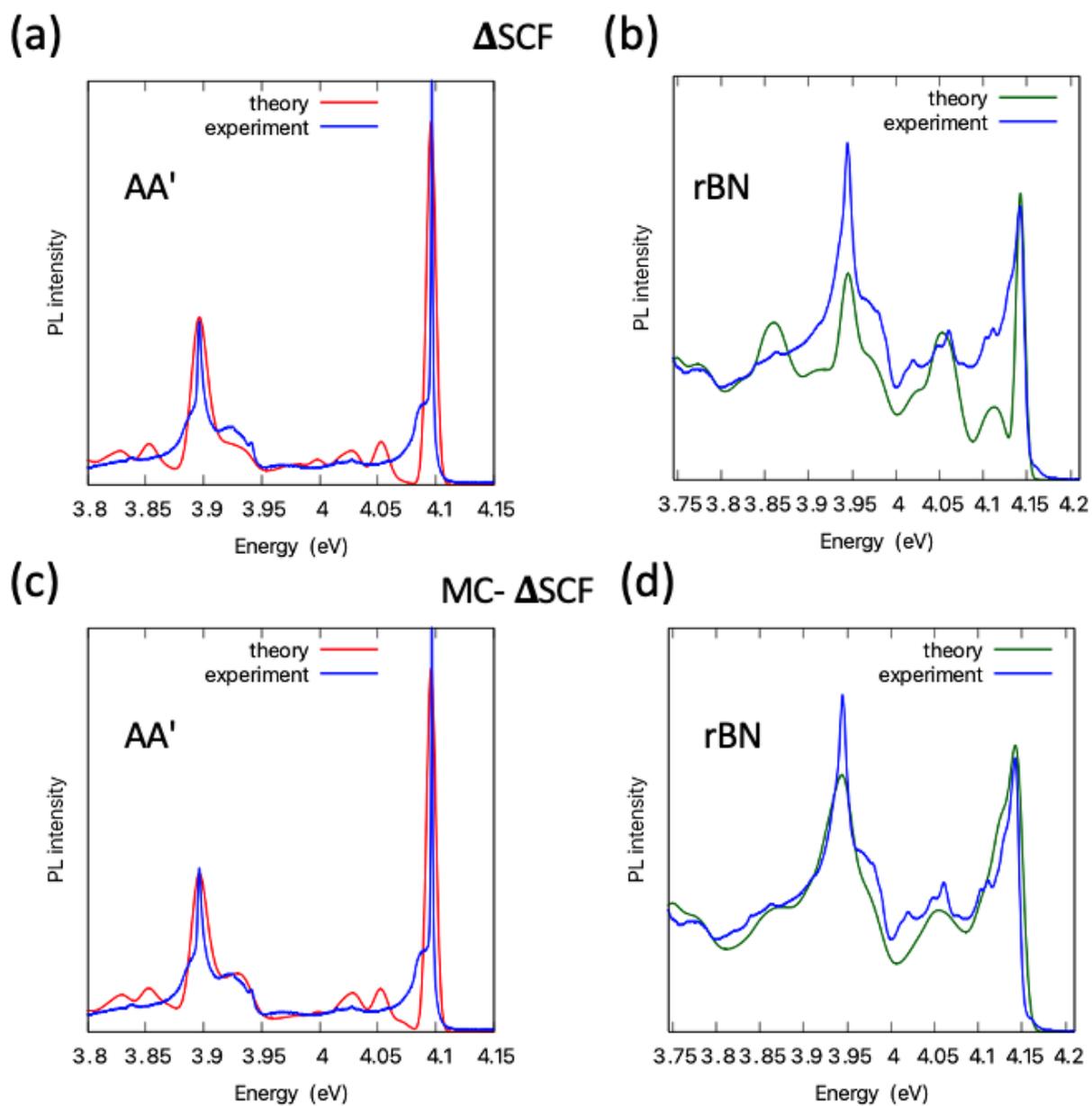

Figure S3. Photoluminiscence spectra calculated for the $C_BC_N$ defect in hBN – AA' (b and d) and rBN - ABC (c and e). The spectra (b) and (c) were obtained with the standard ΔSCF, while in the calculations of (d) and (e), the MC-ΔSCF method was employed. Our experimental results are provided for a sake of reference.

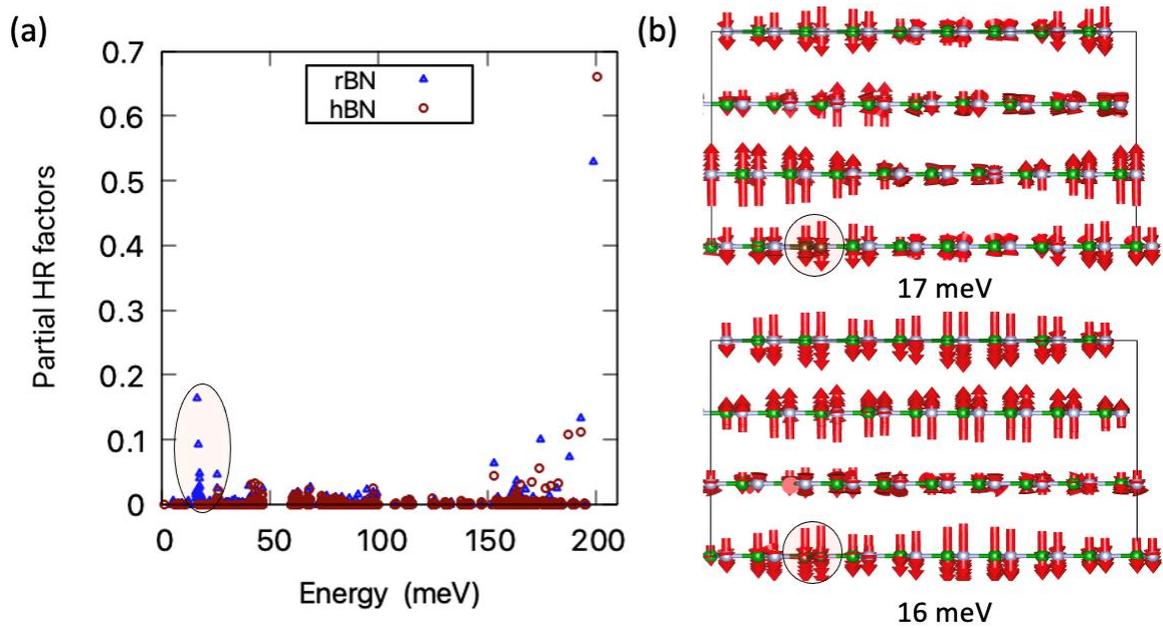

Figure S4. (a) Partial Huang-Rhys factors, calculated for the $C_BC_N$ defect in hBN and rBN. (b) Low-energy out-of-plane modes, active in rBN. The arrows show the atomic contributions to the modes.

## Characterization of samples $S_1$, $S_2$, $S_3$

Scanning electron microscopy pictures of samples $S_1$, $S_2$, and $S_3$ are presented in Figure S5. The morphology of the samples differs significantly from that of the reference rBN samples (see Fig. S1). The surface appears flat and clean, devoid of any crystalline debris or precipitates. Notably, a characteristic wrinkle pattern is observed. These wrinkles form during the cooling process of the material after growth, serving as a means for the layer to relax [3,4]. The angle between the wrinkles at knots measures approximately 120° with minor deviations, consistent with the expected hexagonal symmetry of the crystal structure. Additionally, the direction of the wrinkles is parallel to each other, suggesting high crystalline quality and maintenance of crystal orientation over extended distances.

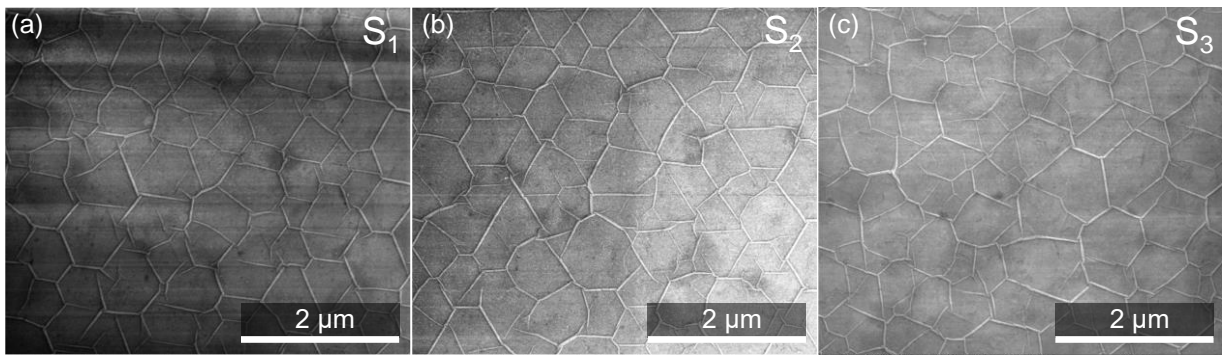

Figure S5: Scanning electron microscopy pictures of samples (a) $S_1$, (b) $S_2$, and (c) $S_3$.

X-ray diffraction results for samples $S_1$, $S_2$, and $S_3$ are illustrated in Figure S6. Similar to the SEM findings, the samples exhibit remarkable similarity. In the broad spectrum (Fig. S6a), all characteristic features for 2θ/ω scans conducted on epitaxial $sp^2$-BN grown on sapphire are evident. Peaks observed at 20.5°, 40°, and 41.75° correspond to the diffraction on (003) and (006) atomic planes in $Al_2O_3$. Additionally, a broad peak centered at 36° indicates the presence of aluminum nitride. This thin AlN layer (typically 1-3 nm) forms at the $Al_2O_3$-BN interface during the initial growth stage and is characteristic of BN grown on sapphire substrates by MOVPE [5]. The peak associated with the atomic interplane distances in $sp^2$-BN appears at 26.5°, providing further insights into the crystal structure of the samples. A magnified view of this angular position is presented in Fig. S6b. The peak exhibits asymmetry, indicating the presence of two components. The more intense component, located around 26.7°, originates from the main part of the material, characterized by high crystalline quality and an intralayer distance very close to the expected 3.33 Å for $sp^2$-BN [5–8]. A broader component of the peak is centered at a lower angular position, attributable to inclusions of lower crystalline quality exhibiting turbostratic behavior [5,9]. Random twisting and sliding of successive atomic layers result in an increased lattice constant in the *c* direction. The interlayer distances determined from fitting two Gaussian curves to the observed peaks for the samples are 3.33 Å and 3.35 Å for $S_1$, 3.33 Å and 3.35 Å for $S_2$, and 3.34 Å and 3.38 Å for $S_3$, corresponding to well-ordered and turbostratic phases, respectively. Although we refer to the less-ordered material as turbostratic, its crystal quality is relatively high compared to epitaxially grown boron nitride [10–13].

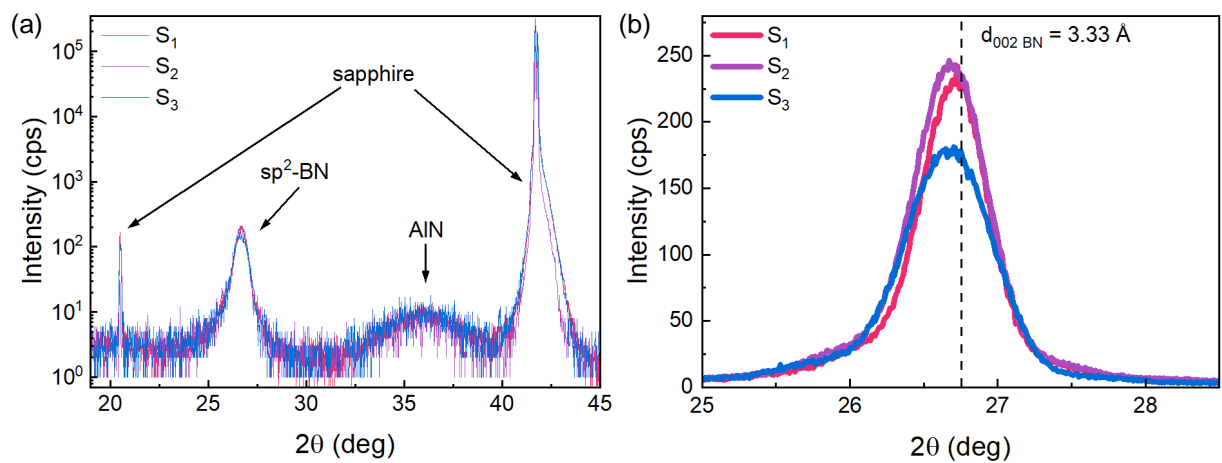

Figure S6: 2θ/ω scans for samples $S_1$, $S_2$, and $S_3$ in a broad angular range (a), and narrow range related to the diffraction on atomic interplane (b). Dashed line indicated at 26.75° angular position refers to the atomic layers distance of 3.33 Å expected for $sp^2$-BN.

Fourier-transform infrared (FTIR) spectra were collected using a Thermo Scientific Nicolet Continuum Infrared Microscope equipped with a 32x Schwarzschild infinity-corrected objective (NA 0.65). Measurements were conducted at five distinct 70×70 μm areas, spaced approximately 5 mm apart, to ensure high homogeneity. High reflection below 1000 cm$^{-1}$, as shown in Figure S7, originates from the sapphire substrate. The peak at approximately 1367 cm$^{-1}$ represents the $E_{1u}$ vibrational mode in sp$^2$-BN [4,14,15]. A magnification of the spectral range associated with this peak does not allow for the observation of significant differences between the samples. Therefore, we employed a fitting procedure that treats the material as a medium composed of damped harmonic oscillators, as described in detail in ref. [4]. The obtained values for the oscillator self-energy are 1367.1 cm$^{-1}$, 1367.2 cm$^{-1}$, and 1367.5 cm$^{-1}$, with damping parameters of 17.4 cm$^{-1}$, 17.6 cm$^{-1}$, and 18.4 cm$^{-1}$, and thicknesses of 19.9 nm, 17.5 nm, and 15.3 nm for samples $S_1$, $S_2$, and $S_3$, respectively. It is evident that a longer growth time at 1400 °C (sample $S_3$) leads to a decrease in growth rate, an increase in strain, and a deterioration of optical quality, as indicated by the thickness, oscillator self-energy, and damping parameter, respectively. The obtained values of the damping parameter, which also correspond to the peak width in FTIR and Raman spectroscopies, are very low compared to the values typically observed in MOVPE-grown sp$^2$-BN.

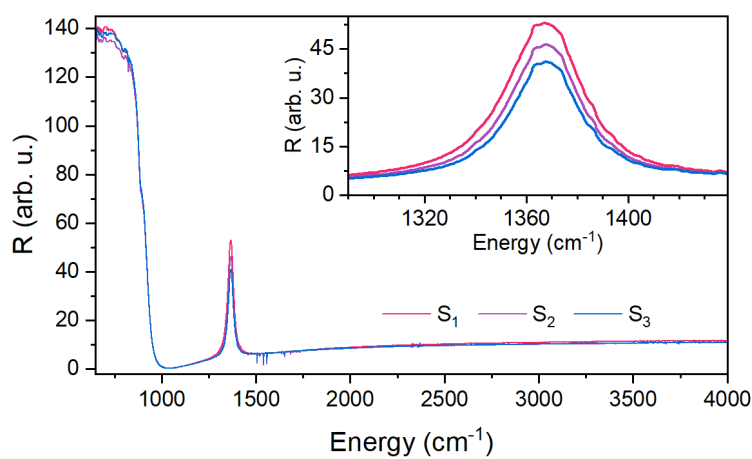

Figure S7: Fourier-transform infrared spectra for samples $S_1$, $S_2$, and $S_3$. Inset shows zoom on the spectral range associated with $E_{1u}$ vibrational mode in sp$^2$-BN.

## Cathodoluminescence fitting procedure

The obtained cathodoluminescence (CL) spectra can be decomposed into five component curves. The measured line-scans of CL spectra for BN layers with heterogeneous polytypic composition allowed us to identify spectra with a single ZPL, which can be attributed to a specific polytype. Thus, from these experimental data points, we extracted components related to pure rBN and hBN. X-ray diffraction analysis indicated the presence of less ordered turbostratic inclusions in our material. As demonstrated in ref. [1], the twisting of layers results in peak broadening. To address this effect, we consider three Gaussian curves positioned at the zero-phonon line and phonon replicas at 3.70 eV, 3.90 eV, 4.09 eV with widths of 0.21 eV, 0.18 eV, 0.13 eV, respectively. The spectral fitting procedure involves adjusting the intensity of each component, while other parameters remain unchanged. Therefore, the fitting of the entire spectrum is simplified to only five parameters. An exemplary spectrum deconvolution into the rBN, hBN, and three Gaussian background components is presented in Figure S8. The intensity of the components presented in Figure 5 in the main text is related to the fitted intensity of rBN and hBN components according to the described procedure. An animation demonstrating the variation in CL spectra along a single line-scan is provided in the additional supplementary materials file. Each spectrum is displayed with all five components and their fitted intensities.

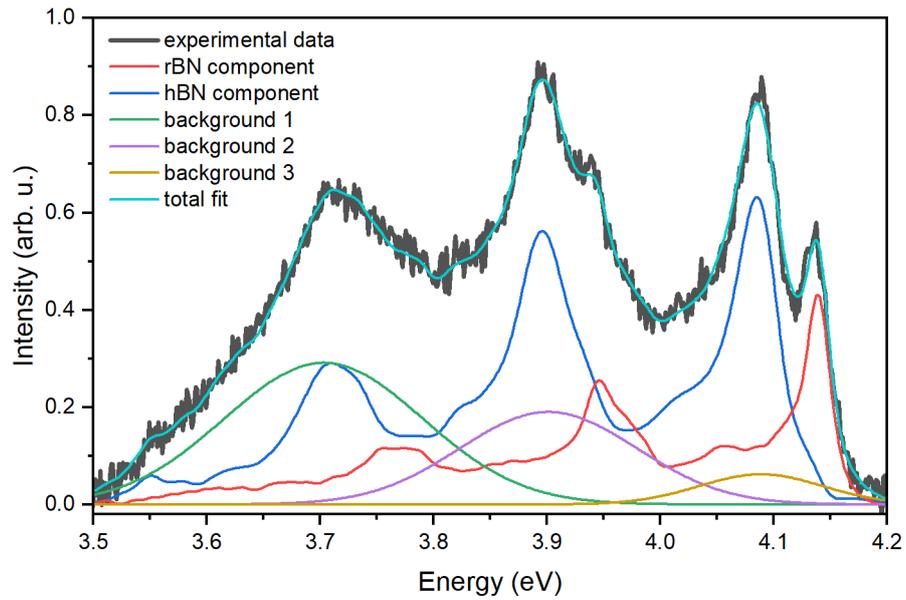

Figure S8: Exemplary cathodoluminescence spectrum for sample $S_1$ (gray line) with fitted curve (cyan line). Red and blue lines represent rBN and hBN spectrum components, respectively. Green, purple and yellow lines correspond to background signal originating from twisting of atomic layers.

## High-resolution transmission electron microscopy

In Figure S9, high-resolution transmission electron microscopy images of samples $S_1$, $S_2$, and $S_3$ are depicted. These images cover a wider area for the figures presented in the main text (Fig. 5a-c). As can be noticed, the $sp^2$-BN layers are oriented parallel to the sapphire substrate and comprise a mixture of different polytypes. The areas presented in the main text are delineated by white frames. It is essential to refrain from drawing conclusions regarding the proportions between individual polytypes based solely on these images. Instead, the composition analysis should rely on multiple lamella photos taken from various parts of the sample. Our statistical analysis conducted in this manner provided evidence of the ratio between individual polytypes, consistent with the results of more global techniques like XRD, CL linescans and macroPL.

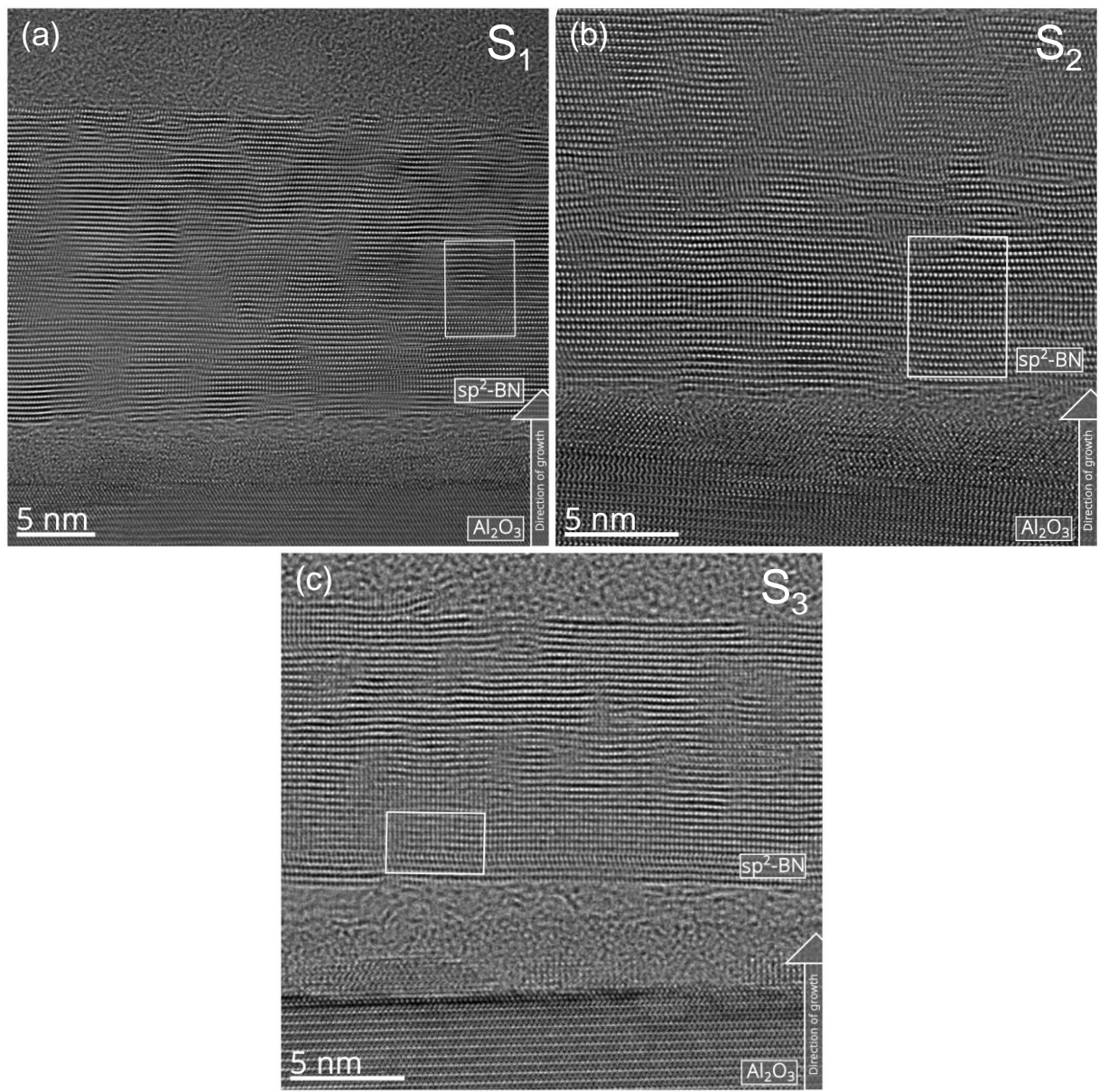

Figure S9: High-resolution transmission electron microscopy pictures of FIB cross-section of the samples (a) S1, (b) S2, (c) S3. White frames show the areas presented in the main text.